\documentclass[12pt]{article}
\usepackage{latexsym}
\usepackage{amsmath}
\usepackage{amssymb}
\usepackage[dvips]{graphicx}
\usepackage{latexsym}
\usepackage{enumerate,graphicx}
\usepackage[toc,page]{appendix}
\usepackage{setspace}
\usepackage{color,soul}
\newcommand{\C}{\mathbb{C}}

\newcommand{\R}{\mathbb{R}}

\newcommand{\koniec}{\begin{flushright}  $\Box $ \end{flushright}}
\def\be{\begin{equation}}
\def\ee{\end{equation}}

\def\t{\tilde}

\def\wh{\widehat}

\def\vv{\varepsilon}

\def\Om{\Omega}

\def\g{\gamma}

\def\p{\partial}

\def\k{\kappa}

\def\ra{\rightarrow}

\def\h{\hat}

\def\om{\omega}
\def\mc{\mathcal}

\def\dsl{\displaystyle}

\newcommand{\hook}{{\setlength{\unitlength}{11pt}   
                   \begin{picture}(.833,.8)
                   \put(.15,.08){\line(1,0){.35}}
                   \put(.5,.08){\line(0,1){.5}}
                   \end{picture}}}
\def\a{\alpha}
\def\ll{\lambda}

\newtheorem{theo}{Theorem}[section] 
\newtheorem{prop}[theo]{Proposition}

\begin{document}
\title{
{\bf Skyrme fields, multi-instantons and \\ the $SU(\infty)$-Toda equation \vskip 20pt}}

\vskip 100pt

\author{
Prim Plansangkate\thanks{
  Email: prim.p@psu.ac.th} \\[15pt]
{\it Applied Analysis Research Unit}, \\
{\it Department of Mathematics and Statistics}, \\
 {\it Prince of Songkla University}, \\
{\it Hat Yai, Songkhla, 90110 Thailand}.}

\date{}
\maketitle

\vskip 50pt

\begin{abstract}
We construct Skyrme fields from holonomy of the spin connection of 
 multi-Taub-NUT instantons  with the centres positioned along a line in $\R^3.$  
 Our family of Skyrme fields includes the Taub-NUT Skyrme field previously constructed by Dunajski.
 However,  we demonstrate that different gauges of the spin connection can result in Skyrme fields with different topological degrees.  
As a by-product, we present a method to compute the  degrees of the Taub-NUT and 
Atiyah-Hitchin Skyrme fields analytically; these degrees are well defined  as  a preferred gauge is fixed by the $SU(2)$ symmetry of the two metrics.

Regardless of the gauge, the domain of our Skyrme fields is the space of orbits of the axial symmetry of the multi-Taub-NUT instantons.  
We obtain an expression for the induced Einstein-Weyl metric
on the space and its associated solution to the $SU(\infty)$-Toda equation.

\end{abstract}

\newpage
\setcounter{equation}{0}

\section{Introduction}

While the attempt to unify gravity into the same framework as that of quantum field theories of particles is still the most
ambitious goal in theoretical physics, a geometric model of particles which uses Riemannian geometry rather than field theory
 has been proposed in \cite{AMS12}.  In this paper, Atiyah, Manton and Schroers (AMS) proposed a description of elementary particles in terms of self-dual 
gravitational instantons.  These are 4-dimensional Riemannian manifolds which are Einstein, having self-dual Weyl tensor, and whose curvature decays at infinity 
\cite{GH79}.   The topological invariants of the manifolds
are identified with quantum numbers of particles.  For example, the electrically charged particles are modelled by non-compact asymptotically 
locally flat (ALF) instantons, where the electric charge is given by the first Chern class of the asymptotic circle fibration.  The baryon 
number is proposed to be the signature of the manifold.  

The AMS model is inspired by an older geometric model of particles, called the Skyrme model \cite{S62}.  This is a nonlinear
scalar field theory in $3$ spatial dimensions, with the Skyrme fields taking values in $SU(2).$  It can be regarded as an effective, low energy approximation of Quantum Chromodynamics in the limit of large number of quarks.
In the Skyrme model, baryons are described by soliton solutions of the theory.   A Skyrme field at a given time is a map
$U: \R^3 \ra SU(2),$ which satisfies the boundary condition $U({\bf x}) \ra {\bf 1}$ as ${\bf x} \ra \infty.$  This boundary condition ensures that
the topological degree of the map $U$ is well defined, and it is then identified with baryon numbers.
Static Skyrme fields are critical points of the energy functional derived from the Skyrme Lagrangian.  The minimum energy configuration in each topological
degree sector is called Skyrmion.
The Euler-Lagrange equation of the Skyrme model is not integrable.   However, good approximations of the minimum energy solutions can be generated 
from $SU(2)$ Yang-Mills instantons in $\R^4$.   This is done by calculating the 
holonomy of the Yang-Mills instantons along lines in a fixed direction \cite{AM89}.    We shall also called the resulting scalar fields Skyrme fields.
  The topological degrees of the Skryme fields are 
given by the instanton numbers of the corresponding Yang-Mills fields.

While it is shown in \cite{AMS12} how gravitational instantons describe some static particles, namely electron, proton, neutron and neutrino, and 
how quantum numbers can be interpreted as topological invariants, yet a relation with the Skyrme model had not been explored.  
This was later done in \cite{D13}, where a construction of a Skyrme field from a holonomy of the self-dual spin connection of a gravitational instanton has been proposed.
The gravitational instanton is assumed to be self-dual and Ricci-flat. 
The construction was applied to the Taub-NUT and Atiyah-Hitchin instantons, which in the AMS model describe electron and proton, respectively.

Motivated by the work in \cite{D13}, this paper aims to explore further implementation of the 
construction in \cite{D13} to a family of gravitational instantons, namely the multi-Taub-NUT instantons \cite{GH79}. 
The multi-Taub-NUT instantons form a family of ALF instantons, known as type $A_{N-1},$  where $N=1$ corresponds to the Taub-NUT instanton.
The multi-Taub-NUT metric is known explicitly as
\be \label{multiTNmet} g = V(dr^2 + r^2 (d\theta^2 + \sin^2 \theta d \phi^2)) + V^{-1}(d \psi + \a)^2, \qquad  V = 1 + \sum_{n=1}^{N} \frac{1}{|| {\bf x} - {\bf x}_n ||}, \ee
where $r \in [0, \infty),\; \theta \in [0, \pi], \; \phi \in [0,2\pi)$ are the usual spherical coordinates, 
 the metric ${dr^2 + r^2 (d\theta^2 + \sin^2 \theta d \phi^2)}$ is the flat metric on $\R^3,$  and  the range  of $\psi$  is $[0, 4 \pi).$ \footnote{This is to remove the Dirac String singularities running from each centre \cite{GH78, GH79}.}  The function $V({\bf x})$ is a positive solution 
to the Laplace's equation and ${\bf x}_n$ denote $N$ distinct points on $\R^3.$ 
The one-form $\a$ is related to $V$ via $d \a = *_3 dV,$ where $*_3$ is the Hodge star operator with respect to the flat metric.

Here, we shall only consider the multi-Taub-NUT instantons where the points ${\bf x}_n,$ the centres, lie along the 
union of the lines $\theta =0$ and $\theta = \pi,$ i.e. the $z$-axis in the Cartesian coordinate system $(x,y,z)$ for $\R^3.$
Such a multi-Taub-NUT metric has two generators of Killing symmetry, one is the vector field $\dsl \frac{\p}{\p \psi}$ and the other 
$\dsl \frac{\p}{\p \phi}.$  We shall only use the Killing vector field $\dsl \frac{\p}{\p \phi}$ for the Skyrme field 
construction, as it gives rise to a Skyrme field with nonzero
topological degree in the case of the Taub-NUT instanton \cite{D13}. 

\vspace{0.3cm}

Following the construction in \cite{D13}, we shall obtain the $SU(2)$ Yang-Mills connection on the multi-Taub-NUT background from the 
the spin connection of the multi-Taub-NUT metric  itself.  Such a procedure was first introduced in \cite{CD771}, and further investigated in 
\cite{CD772, PY78, OPY11}.   The holonomy of the Yang-Mills connection will then be calculated along the orbits of the axial  
symmetry of the multi-Taub-NUT instanton. This will give rise to an $SU(2)$-valued scalar field $U$ - our Skyrme field  - on the space of orbits of the 
Killing symmetry.    

In the next section we shall review the construction  \cite{D13} of Skyrme fields from gravitational instantons,
and apply it to the multi-Taub-NUT instantons.  We choose the frame fields for the self-dual spin connection to be a natural extension of the  
$SU(2)$-invariant frame fields used in  \cite{D13}.   We then obtain an explicit expression for the multi-Taub-NUT Skyrme fields, in term of $V$ and $\a$ in (\ref{multiTNmet}); 
the family includes the Taub-NUT Skyrme field  previously constructed in \cite{D13} as the case $N=1.$
In Section \ref{sec:topdeg} we shall however show  that different gauges of the spin connection can result in Skyrme fields with unequal topological degrees.  
In particular, writing the multi-Taub-NUT spin connection in a different set of frame fields,  we show that the resulting family of Skyrme fields  has 
vanishing  topological degree for all $N \ge 1.$   Unlike the Taub-NUT and 
Atiyah-Hitchin Skyrme fields where the $SU(2)$ symmetry is present, it is not obvious how one can justify a preferred gauge for the multi-Taub-NUT $N\ge 2$ case.  
Nevertheless, as a by-product, 
we present a method to compute the topological degrees of the Taub-NUT and Atiyah-Hitchin Skyrme fields analytically.  
(These degrees are well defined as they are fixed by the frame fields chosen to respect the $SU(2)$ 
symmetry of both metrics.)

The space where the Skyrme fields live is the space of orbits of the axial symmetry of the multi-Taub-NUT metric.  
In Section \ref{sec:SUinf} we investigate the Einstein-Weyl metric on the space. In fact we study a more general case of the 
 Einstein-Weyl metric on the space of orbits  of a  Killing 
vector field of the form $\dsl \frac{\p}{\p \phi} + c \frac{\p}{\p \psi},$ where $c$ is a constant. In particular, we obtain implicit expressions for the metric and its
associated solution to the $SU(\infty)$-Toda equation.   
Lastly, in Section \ref{sec:Discuss}, we discuss the Skyrme energy functional, particularly for the
Taub-NUT Skyrme field, and remark how one could proceed to compare the energy of the approximate Skyrme field with that of the true solution to the Euler-Lagrange  equation.

\vspace{1cm}


\section{Skyrme field construction}  \label{sec:Skyrmconstr}

The Skyrme field construction of \cite{D13} is based on two important results.  One is that  a solution of the self-dual Yang-Mills equation on 
a Ricci-flat background can be obtained from the spin connection of the background metric itself \cite{CD771}. The other is that static Skyrme fields
can be generated from $SU(2)$ Yang-Mills instantons \cite{AM89}.

\subsection{From spin connection to Yang-Mills instanton}

Let $(M, g)$ be a Riemannian spin four manifold with a Ricci-flat metric $g,$ 
 and $\om_{ab}$ be the connection one-form defined from an 
orthonormal tetrad of one-forms $\{ e_a \}$ by 
\[ d e_a = \om_{ab} \wedge e_b,  \quad \om_{(ab)} = 0, \quad a, b = 0,1,2,3. \]
 Charap and Duff \cite{CD771} showed that one can identify $\om^{ab}$ with a $Spin(4)$ Yang-Mills potential, which under the decomposition
$Spin(4) = SU(2) \times SU(2)$
gives rise to an $SU(2)$ Yang-Mills potential satisfying the (anti-) self-dual equation on the metric background.

Recall (see, for example, \cite{D09}) that the complexified tangent bundle can be decomposed as $TM \otimes \C \cong S_+ \otimes S_-,$ where $S_{\pm}$ are rank two complex
vector bundles.  The spin bundles $S_{\pm}$ inherit connections $\g_\pm$ (spin connections) from the Levi-Civita connection of $g$. 
 Let us choose a convention such that  $\g_+$ encodes the information
about the self-dual part of the Weyl tensor, and $\g_-$ the anti-self-dual part.  Then it means that given a self-dual Ricci-flat 
metric, one can interpret the spin connection $\g_+$ as an $SU(2)$ self-dual Yang-Mills connection on the metric background.

Pope and Yuille applied such a procedure to construct an  $SU(2)$ self-dual Yang-Mills instanton in the Taub-NUT background \cite{PY78}. The result was then used in
one of the examples in \cite{D13} to generate a Skyrme field from a gravitational instanton, namely the Taub-NUT Skyrme field.   The goal of our paper
is to extend the Skyrme field construction in  \cite{D13} to multi-Taub-NUT instantons.  Let us now describe how
one can define a Yang-Mills instanton from the spin connection of a multi-Taub-NUT metric.

The multi-Taub-NUT metrics form a family of gravitational instantons which are hyperK\"ahler, thus they are Ricci-flat and have self-dual
 Riemann tensor.\footnote{The standard convention for a hyperK\"ahler manifold is to choose an orientation determined by the square of one of the K\"ahler forms.  With respect to this orientation the K\"ahler forms are self-dual while the Weyl tensor is anti-self-dual.  However, we have chosen the reverse orientation, where the Weyl tensor is self-dual, 
in order to facilitate the readers in comparing the results with those in previous literature such as \cite{PY78, AMS12, D13}.    }  
They are asymptotically flat (ALF), meaning the metrics approach $S^1$ bundles over $S^2$ at infinity.  The metrics of the multi-Taub-NUT family, 
also called the type $A_{N-1},$ are given explicitly by the Gibbons-Hawking ansatz \cite{GH78} (\ref{multiTNmet}).  They have a triholomorphic $S^1$ symmetry, generated by 
the vector field $\dsl \frac{\p}{\p \psi}.$ \; For the purpose of our Skyrme field construction, we shall only consider a subclass of the 
multi-Taub-NUT instantons, where the points ${\bf x}_n,$ the centres, lie along the $z$-axis in the Cartesian coordinate system.  
This results in the function $V$ being independent of $\phi.$ 

The self-dual spin connection, $\g:=\g_+,$ of a multi-Taub-NUT metric can be 
calculated using the self-dual two-forms
\be \label{sd2form} \Sigma_i = e_0 \wedge e_i + \frac{1}{2} \vv_{ijk}  e_i \wedge e_j, \qquad  i, j, k = 1,2,3,\ee
where $\{e_0, e_i\}$ is an orthonormal tetrad of one-forms.  The spin connection coefficients, $\g_{ij},$ are
determined from 
\be \label{Cartan} d \Sigma_i + \g_{ij} \wedge \Sigma_j =0.\ee

Consider the metric of the form 
in (\ref{multiTNmet})
\be \label{multiTNsph} g = V(dr^2 + r^2 (d\theta^2 + \sin^2 \theta d \phi^2)) + V^{-1}(d \psi + \a)^2, \ee
where $V$ is now independent of $\phi$ and is given by
\be \label{Vsph} V(r, \theta) = 1 + \sum_{n=1}^{N} \frac{1}{\sqrt{r^2 - 2 z_n r \cos \theta + z_n^2}} \, , \ee
and the one-form $\a$ satisfying $*_3 d V = d \a$ is chosen to be
\be \label{asph} \a  = \h \a(r, \theta) \, d \phi, \qquad \mbox{where} \quad \h \a(r, \theta) = \sum_{n=1}^{N} \frac{r \cos \theta - z_n}{\sqrt{r^2 - 2 z_n r \cos \theta + z_n^2}}\, . \ee

\vspace{0.3cm}

Now, let
\vspace{-1cm}
\begin{eqnarray}
\eta_1 &=& -\sin \psi \, d \theta + \sin \theta \cos \psi \, d \phi,  \nonumber \\  
 \label{eta} \eta_2 &=& \cos \psi \, d \theta + \sin \theta \sin \psi \, d \phi, \\
\eta_3 &=& d \psi + \h \a \, d \phi. \nonumber
\end{eqnarray}
Then (\ref{multiTNsph}) becomes
\[ g = V (d r^2 + r^2 (\eta_1^2 + \eta_2^2))  + V^{-1} \eta_3^2.\]

We can then choose an orthonormal tetrad of one-forms  to be
\be \label{sphtetrad} e_0 = \sqrt{V} \, d r, \quad e_1 = r \sqrt{V} \, \eta_1, \quad e_2 = r \sqrt{V} \, \eta_2, \quad e_3 = \frac{1}{\sqrt{V}} \, \eta_3.  \ee

\vspace{0.3cm}

For the Taub-NUT metric, with $N=1, z_1 =0,$ we have that $\dsl V= 1+\frac{1}{r},$ $\h \a = \cos \theta$ and $(\eta_1,\eta_2, \eta_3)$ in (\ref{eta})
are left-invariant one-forms on $SU(2),$  which satisfy
\[ \label{etarelationTN} d \eta_1  = \eta_2 \wedge \eta_3, \qquad d \eta_2  = \eta_3 \wedge \eta_1, \qquad d \eta_3  = \eta_1 \wedge \eta_2.  \]
Thus the tetrad (\ref{sphtetrad}) is $SU(2)$-left invariant.

\vspace{0.3cm}

For the multi-Taub-NUT instanton, the metric no longer has spherical symmetry and $(\eta_1, \eta_2, \eta_3)$ in (\ref{eta}) are not left-invariant one-forms
on $SU(2).$  However, to obtain a family of Skyrme fields which includes the Taub-NUT Skyrme field in \cite{D13}, 
we shall proceed by using (\ref{sphtetrad}) as our orthonormal tetrad, with $V$  and $\a$ given  in (\ref{Vsph}),(\ref{asph}).

 The spin connection coefficients, $\g_{ij},$ are
determined from (\ref{Cartan}).  Then the  $SU(2)$ Yang-Mills potential, $A$,  is given by
\be \label{YMpotential} A = \frac{1}{2} \vv_{ijk} \,  \g_{jk} \otimes {\bf t}_i,\ee
where $\{{\bf t}_i\}$ are generators of the Lie algebra ${\bf su}(2)$ satisfying 
$\dsl [{\bf t}_i, {\bf t}_j] = - \vv_{ijk} {\bf t}_k.$

\vspace{0.3cm}

That is, 
\[ A=A_1 \otimes {\bf t}_1 + A_2 \otimes {\bf t}_2 +A_3 \otimes {\bf t}_3, \]
where $A_{1} = \g_{23}, \, A_2 = \g_{31}$ and $A_3 = \g_{12}.$    We find that

\begin{eqnarray} \label{gamma}
A_1 &=& \frac{\sin \psi}{r V \sin \theta} \, \h \a_r \, dr - \left(  1+ \frac{r V_r}{V} \right) \, \eta_1 - \frac{\cos \psi}{r V^2 \sin \theta} \, \h \a_r \, \eta_3  \nonumber \\  
A_2 &=& - \frac{\cos \psi}{r V \sin \theta} \, \h \a_r \, dr - \left( 1+ \frac{rV_r}{V} \right) \, \eta_2 - \frac{\sin \psi}{r V^2 \sin \theta} \, \h \a_r \, \eta_3 \\
A_3 &=& -\frac{\cos \psi}{V \sin \theta} ( \h \a_r+ V (\h \a-\cos \theta) ) \, \eta_1  - \frac{\sin \psi}{V \sin \theta}  ( \h \a_r+ V (\h \a-\cos \theta) ) \, \eta_2
+ \left( 1+ \frac{V_r}{V^2}  \right) \, \eta_3, \nonumber
\end{eqnarray}
which gives the result in \cite{D13} for the Taub-NUT case when $\h \a =\cos \theta$ and $\dsl V= 1+\frac{1}{r}.$


\subsection{From Yang-Mills instantons to Skyrme fields}

Atiyah and Manton showed in \cite{AM89} that Skyrme field configurations in $\R^3$ can be generated from $SU(2)$
Yang-Mills instantons.  Given any such Yang-Mills field in $\R^4,$  the holonomy of the Yang-Mills field along all lines
parallel to the time axis gives rise to a scalar $SU(2)$-valued function $U$ on $\R^3.$  It was shown that $U$ 
 satisfies the boundary condition $U({\bf x}) \ra {\bf 1}$ as ${\bf x} \ra \infty$ and thus can be regarded as a Skyrme field.
Dunajski \cite{D13} applied the Atiyah-Manton construction to the Yang-Mills fields from gravitational instantons, namely
the Taub-NUT and the Atiyah-Hitchin instantons.  One difference is that the Yang-Mills instantons now live on  curved backgrounds.  
It is noted that the holonomy should be calculated along orbits of a Killing vector field of the
background manifold, so that the space of orbits $\mc{B},$  where the Skyrmion lives, admits a metric.  

Suppose $K$ is a Killing vector field and $x^a = (s, {\bf x})$ be coordinates on the manifold $M$ such that
$\dsl K = K^a \frac{\p}{\p x^a} = \frac{\p}{\p s}.$  \, Then formally the holonomy is given by
\be \label{holonomy} U = \mc{P} \exp \left(- \int_\Gamma A \right), \ee
where $\mc{P}$ denotes $s$-ordering.    In general, one computes $U$ by
solving the equation  ${K^a D_a \Psi = 0,}$ where $D_a = \p_a +A_a.$  For a non-compact $\Gamma,$ one
solves for a scalar $SU(2)$-valued function $\Psi(s, {\bf x})$ under the initial condition  $\Psi(-\infty, {\bf x}) = {\bf 1}.$  
 Then $U({\bf x}) =  \Psi(\infty, {\bf x}).$  If the orbit $\Gamma$ is $S^1,$ one needs to cut it into an interval.
In \cite{D13}, the holonomy is calculated along the $S^1$ orbits of a left translation $SO(2)$ inside $SU(2),$ generated by the vector field $\dsl \frac{\p}{\p \phi}.$
The orbits generated by 
 $\dsl \frac{\p}{\p \psi}$ was not considered as it gives an Abelian Skyrme field with zero topological degree.

\vspace{0.5cm}

We shall now follow \cite{D13} and choose the orbit $\Gamma$ to be that generated by $\dsl \frac{\p}{\p \phi},$ which is a Killing 
vector field for our axially-symmetric multi-Taub-NUT instantons.  Let the component $A_\phi$ be the restriction of $A$ to the orbit
$\Gamma.$  Thus,
\be \label{integralUphi} U = \mc{P} \exp \left(- \int_0^{2\pi} A_\phi \, d \phi\right). \ee

As we shall see in Section \ref{gdepend} that this construction of Skyrme fields is gauge-dependent, so let us first discuss our gauge choice of the Yang-Mills potential $A$ in (\ref{gamma}).

The gauge is chosen as a generalisation of the one used in \cite{D13} so that  the family of the resulting skyrme fields will include the Taub-NUT skyrme field already constructed there.
In the Taub-NUT case, $N=1$ with the centre located at $r=0$, it can be verified by taking the limit $r \ra 0$ that $A$ is regular at the centre.  (The author of \cite{D13} fixed the gauge by demanding
the regularity at $r=0.$)  However, for the multi-Taub-NUT instanton with $N \ge 2,$ the potential $A$  has apparent singularities at the centres. This is a gauge phenomenon, which can be seen by considering the Yang-Mills curvature $F$ of  $A$ in (\ref{YMpotential})
\[ F = dA + A \wedge A = \frac{1}{2} \vv_{ijk} R_{jk} \otimes {\bf t}_i, \]
where $R_{ij} = d \g_{ij} + \g_{ik} \wedge \g_{kj}$ is the Riemann curvature two-form of the multi-Taub-NUT metric which is regular at the centres.

It is therefore possible to remove the apparent singularities of $A$ in gauge (\ref{gamma}) by gauge transformations (see \cite{GT03, GS11}), but this would change the family of Skyrme fields.  Instead, we shall argue that the integrand of 
the path-ordered integral (\ref{integralUphi}) nevertheless remains finite as one approaches a centre.

To see this, let $A_\phi = \g_1 \otimes {\bf t}_1 + \g_2 \otimes {\bf t}_2 + \g_3 \otimes {\bf t}_3,$ where 
$\dsl \g_j = \frac{\p}{\p \phi} \hook A_{j}.$   \\
 From (\ref{eta}) and (\ref{gamma})    we have
\begin{eqnarray} \label{gammaphi}
\g_{1} &=& -\cos\psi \left( \sin \theta +\frac{1}{r} \left( \frac{\h \a}{V} \right)_\theta \right) \nonumber \\
\g_{2} &=& -\sin \psi \left( \sin \theta +\frac{1}{r} \left( \frac{\h \a}{V} \right)_\theta \right) \\
\g_{3} &=& \cos \theta - \left( \frac{\h \a}{V} \right)_r , \nonumber
\end{eqnarray} 
where we have simplified the expressions using the relation $*_3 d V = d \a$ in spherical coordinates:
\[ \h \a_r = - \sin \theta \, V_\theta, \qquad \h \a_\theta = r^2 \sin \theta \, V_r.  \] 


Now as we can always choose coordinates so that the origin is at any centre, without loss of generality we consider a centre located at the origin in $\R^3$. 
 The behaviour of $A_\phi$ near the centre can be seen from 
the limit of  $\g_j$ (\ref{gammaphi}) as $r \ra 0.$  By direct calculation, it can be shown that  as $r \ra 0,$  we have $\g_1, \g_2 \ra 0$ and $\g_3$ approaches an integer 
depending on the numbers of the centres above and below the origin along the axis of symmetry.  To be precise, if the position of each centre is determined by the real number $z_n$
in (\ref{Vsph}), setting $z_1=0,$ we have that $\dsl \g_3 \ra \sum_{n=2}^{N} \frac{z_n}{|z_n|},$ which is finite.

\vspace{0.5cm}

Let us now evaluate the path-ordered integral (\ref{integralUphi}).
Since all the spin connection coefficients are independent of $\phi,$ the integral (\ref{integralUphi}) can be evaluated 
explicitly.  This gives
\[U = \exp \left(-2\pi A_\phi \right), \]
where $A_\phi = \g_1 \otimes {\bf t}_1 + \g_2 \otimes {\bf t}_2 + \g_3 \otimes {\bf t}_3.$ 

Suppose we choose the representation $\dsl {\bf t}_j = \frac{i}{2} \tau_j,$ where $ \tau_j$ are 
Pauli matrices.  Then the Skyrme field from the axially-symmetric multi-Taub-NUT instanton is given by
\be \label{UmultiTN} U(r, \theta, \psi) = \exp \left( -i \,\pi \, \g_{j} \, \tau_j\right), \ee
where $\g_{j }$ are given in (\ref{gammaphi}).  The expression (\ref{UmultiTN}) simplifies in the case
of the Taub-NUT instanton (see \cite{D13}).   \, Also, using (\ref{gammaphi}) it can be shown that 
 the multi-Taub-NUT Skyrmion approaches a constant group element at infinity.  In particular, like the Taub-NUT case, 
 $U \ra {- \bf 1}$ as $r \ra \infty.$

The Skyrme field generated is defined on the space of orbits of the $\dsl \frac{\p}{\p \phi}$ symmetry.
We shall denote this space as ${\mc B}.$ The Einstein-Weyl metric on ${\mc B}$ is investigated in Chapter \ref{sec:SUinf}, with relation to the $SU(\infty)$-Toda equation discussed.

We also note here that the $S^1$ orbits along which we calculate the holonomy have 
no point in common.  Therefore, unlike in the Atiyah-Manton construction \cite{AM89}, the initial condition at the based point
 of all the orbits cannot be imposed. Thus gauge transformation of the Yang-Mills potential (\ref{YMpotential}) will affect the resulting Skyrme field.  
To have a well defined Skyrme field, a preferred gauge has to be choosen.   In \cite{D13}, the preferred gauge for the Taub-NUT and Atiyah-Hitchin 
instantons has been fixed by choosing the $SU(2)$ invariant frame fields (\ref{sphtetrad}).  This choice of gauge is resulted from the symmetry 
requirement, that the Yang-Mills potential is $SU(2)$ left-invariant, and the requirement that the potential is regular at $r=0.$
The multi-Taub-NUT metrics for $N \ge 2$ do not have the $SU(2)$ symmetry, and it is not obvious how one can justify a preferred gauge.  However, 
the family of Skyrme fields (\ref{UmultiTN}) can be considered as a generalisation of the Taub-NUT Skyrme field of \cite{D13}, as it includes the Skyrme field in the family.

\vspace{1cm}


\section{The topological degree}  \label{sec:topdeg}

The Skyrme field constructed in Section \ref{sec:Skyrmconstr} is a map from the space of orbits 
${\mc B}$ of the Killing vector field $\dsl \frac{\p}{\p \phi}$ to the Lie group $SU(2).$
An integral expression for a topological degree of the map $U:{\mc B} \ra SU(2)$ is given by
\be \label{degree}  D = - \frac{1}{24 \pi^2} \int_{\mc B} \mbox{Tr} ((U^{-1} d U)^3).\ee

In \cite{D13}, the topological degrees  of the Taub-NUT and Atiyah-Hitchin Skyrme fields are calculated.  It is hoped that they can be interpreted as some physical quantum numbers of the particles
modelled by the gravitational instantons, namely the electron and proton.  Recall that the Taub-NUT and Atiyah-Hitchin Skyrme fields are constructed in the 
gauge preferred by the $SU(2)$ symmetry, and their topological degrees are $2$ and $1,$ respectively.   

However, we shall show below that there exists a tetrad 
of frame fields different from (\ref{sphtetrad}) such that the resulting family of multi-Taub-NUT Skyrme fields has vanishing topological degree for all $N \ge 1.$
  This finding does not present a problem to the physical interpretation of the topological degrees of the 
 Taub-NUT ($N=1$) and Atiyah-Hitchin Skyrme fields, as the symmetry requirement fixes a preferred gauge in which
 the degrees are nonzero.  However, the situation is different for the multi-Taub-NUT case.  Although the choice of frame fields (\ref{sphtetrad}) 
resulting in our multi-Taub-NUT Skyrme fields (\ref{UmultiTN}) is supported by the demand that the family should include  
the Taub-NUT Skyrme field of \cite{D13}, it is not clear to the author if there is a symmetry or regularity requirement that would lead to the gauge
when $N \ge 2.$

As a by-product of this study, we shall also present an analytic method to compute the topological degrees of the Taub-NUT and Atiyah-Hitchin Skyrme fields, 
which were previously obtained in \cite{D13} by the method of preimage counting.

  \subsection{Gauge dependence} \label{gdepend}

Here we shall demonstrate that there exists an orthonormal tetrad of one-forms such that the multi-Taub-NUT spin connection
gives rise to the Skyrme fields with zero topological degree for all $N \ge 1.$ 
Since the multi-Taub-NUT instantons (\ref{multiTNsph}) have axial symmetry, instead of (\ref{sphtetrad}) one could think of another tetrad of one-forms 
which is perhaps more natural with 
regards to the isometry of the metrics.  To define this tetrad, let  us write the metric (\ref{multiTNsph}) in the 
cylindrical coordinates on $\R^3$:  
\[
g = V(d \rho^2 + \rho^2 d\phi^2 + d z^2) + V^{-1}(d\psi +\a)^2,
\]
where now 
\begin{eqnarray*}
V &=& 1 + \sum_{n=1}^N \frac{1}{\sqrt{\rho^2 + (z-z_n)^2}}, \\
 \a  &=& \h \a(\rho, z) \, d \phi, \qquad \mbox{where} \quad \h \a = \sum_{n=1}^N \frac{z-z_n}{\sqrt{\rho^2 + (z-z_n)^2}}.
\end{eqnarray*}

Then a natural tetrad of one-forms to the metric is given by
\be \label{cylintetrad} e_0' = \frac{1}{\sqrt{V}} \, (d \psi + \cos \theta \, d \phi), \quad e_1' = \sqrt{V} \, d \rho, 
\quad e_2' = \rho \sqrt{V} \, d \phi, \quad e_3' = \sqrt{V} \, dz.  \ee

Using this tetrad and repeating the same procedure in  Section \ref{sec:Skyrmconstr} yields a new Yang-Mills potential $A'$  given by
\begin{eqnarray} \label{YMcylin} 
A_1' &=&   \left(\frac{1}{V}\right)_\rho d \psi + \left(\frac{\h \a}{V}\right)_\rho \, d \phi, \nonumber \\  
A_2' &=&  \frac{V_z}{V} \, d \rho  -\frac{V_\rho}{V} \, d z\\
A_3' &=& \left(\frac{1}{V}\right)_z d \psi + \left(\left(\frac{\h \a}{V}\right)_z -1\right) \, d \phi. \nonumber
\end{eqnarray}

Unlike the Yang-Mills field (\ref{gamma}) defined in Section \ref{sec:Skyrmconstr}, the potential (\ref{YMcylin}) 
is independent of the $\psi$ coordinate.  Thus the resulting Skyrme field $U$ only depends on two coordinates of the
three-dimensional space ${\mc B}.$ Therefore the three-form $(U^{-1} d U)^3$ in the integral (\ref{degree}) vanishes and 
gives zero topological degree for all parameter $N \ge 1$. 

 It is well known that under the change of tetrads, the spin connection $\g_{ij}$ 
and thus the Yang-Mills field $A$ change under the usual gauge transformation.  Hence, this demonstrates 
that different gauges of the spin connection result in Skyrme fields with different topological degrees. This presents a problem
to the interpretation of the topological degree as a quantum number of the system of $N$-electrons for $N \ge 2.$


\subsection{Taub-NUT and Atiyah-Hitchin Skyrme fields} \label{sec:TNAH}

We shall end Section \ref{sec:topdeg} by presenting an analytic method to compute the topological degree (\ref{degree}) 
of the Taub-NUT and Atiyah-Hitchin Skyrme fields, constructed in the gauge preferred by the $SU(2)$ symmetry.

First, one notes that the 
integrand of (\ref{degree}) is actually the pullback of the normalised volume form on $S^3 = SU(2).$  
That is, suppose  $(\Theta, \Phi,  \Psi)$ are the Euler angles such that  a point ${\bf q} \in S^3$ can be parametrised by
\be \label{qS3} {\bf q} = \cos \Psi \;{\bf 1} - i \, \bigl(\sin \Psi \sin \Theta \cos \Phi\;  \tau_1\, + \, \sin \Psi \sin \Theta \sin \Phi\;  \tau_2 \, +\, \sin \Psi \cos \Theta \;  \tau_3  \bigr),\ee
where $\Theta \in [0, \pi),$  $\Phi \in [0, 2\pi),$ $\Psi \in [0, \pi)$ and $\tau_j$ are Pauli matrices. 
\,Then 
\[\Om = \sin^2 \Psi \sin \Theta \; d \Psi \wedge d \Theta \wedge d \Phi\]
 is a volume form on $S^3,$  and it can be shown that (\ref{degree})
is given by
\[ D = - \frac{1}{2 \pi^2} \int_{\mc B} { }^*\Om,\]
where ${ }^*\Om$ denotes the pullback of $\Om$ onto ${\mc B}.$ \;
Now, since $\Om$ is closed, it can be written locally as  $\Om = d \h \om$ for some two-form $\h \om$ on $S^3.$  
Moreover, as the exterior derivative $d$ commutes with the pullback, by letting $\om :={ }^*\h \om$ one has that
\be \label{Dom} D =  -\frac{1}{2\pi^2} \int_{\mathcal B}  d \om. \ee
The integral (\ref{Dom}) can be evaluated using the asymptotic values of $\om$ as follows.  

\vspace{0.5cm}

Let $\om = \om_1 \, dr \wedge d \psi +\om_2 \, d\theta \wedge d r +\om_3 \, d\theta \wedge d \psi$ for some functions
$\om_j, \, j=1,2,3.$ \, Then
\be \label{asympint}
\int_{\mathcal B} d \om = \int  \p_\theta \om_1\; d \theta  dr  d\psi +  \int \p_\psi \om_2\; d\psi  d\theta  d r + 
 \int \p_r \om_3\; dr  d\theta  d\psi,    \ee
with appropriate boundaries for the integrals.

Hence, the problem comes down to finding $\om$ and evaluating the relevant asymptotic limits of $\om_j$.    
A natural choice for $\h \om$ is
\be \label{1sthom}\h \om =  \sin \Psi  \cos \Theta \; d \Psi \wedge d \Phi.\ee
Now, to obtain the pullback $\om = {}^* \h \om,$ one needs to find the relation between coordinates
$(\Theta, \Phi,  \Psi)$ on $S^3$ and $(r, \theta, \psi)$ on ${\mc B}.$  

\vspace{0.3cm}

First we note that the Skyrme field $U$ takes value in $SU(2).$  
Now an element in $SU(2)$ of the form  $\exp(-i\, a_j \tau_j)$ for some $a_j$  can be written as 
\[ \exp(-i\, a_j  \tau_j)  =(\cos a) \,{\bf 1} - i (\sin a) \, \h n_j \tau_j,   \]
where  $a_j = a \,\h n_j$ and ${\bf \h n} = (\h n_j)$ is a unit vector.   
In our case, $a_j = \pi \, \g_{j}.$   The Taub-NUT and Atiyah-Hitchin Skyrme fields \cite{D13}  are of the form
(\ref{UmultiTN}), with 
\[\g_1 = f_1(r) \sin \theta \cos \psi, \quad \g_2 = f_2(r) \sin \theta \sin \psi, \quad \g_3 = f_3(r) \cos \theta,\]
for some functions $f_j(r).$ \; Thus, $\dsl \h n_j = \frac{f_j}{\kappa} n_j,$ where 
$\kappa = \sqrt{(f_1 n_1)^2+(f_2 n_2)^2+(f_3 n_3)^2}$  and  ${\bf n} = ( \sin \theta \cos \psi, \sin \theta \sin \psi, \cos \theta).$  Therefore
\be \label{Utau} U = \cos(\pi \kappa) \,{\bf 1} - i \, \sin(\pi \kappa) \, \frac{f_j}{\kappa} n_j \tau_j.  \ee
 
Comparing (\ref{qS3}) and (\ref{Utau}), one concludes that
\begin{eqnarray*}
\cos \Psi &=& \cos (\pi \kappa), \\
\sin \Psi \cos \Theta &=& \frac{\sin (\pi \kappa)}{\kappa} f_3   \cos \theta, \\
\sin \Psi \sin \Theta \cos \Phi &=& \frac{\sin (\pi \kappa)}{\kappa} f_1 \sin \theta  \cos \psi, \\
\sin \Psi \sin \Theta \sin \Phi &=& \frac{\sin (\pi \kappa)}{\kappa} f_2 \sin \theta \sin \psi.
\end{eqnarray*}

\vspace{0.2cm}

From the above relations, it follows that the pullback $\om = {}^*\h \om,$ with $\h \om$ in (\ref{1sthom}), is given by 
\be  \label{1stom}  \om = \frac{\pi \sin^2(\pi \k)}{\k} f_3 n_3 \;   \frac{f_1^2 n_1^2}{f_1^2 n_1^2 +f_2^2 n_2^2} \; d\k \wedge d \left( \frac{f_2}{f_1} \tan \psi \right).\ee

\vspace{0.3cm}

Finally, writing $\om = \om_1 \, dr \wedge d \psi +\om_2 \, d\theta \wedge d r +\om_3 \, d\theta \wedge d \psi$ yields
\[ \om_j =  \beta \, \mu_j,  \]
where
\vspace{-0.4cm}
 \begin{eqnarray*}
\beta &=&  \frac{\pi \sin^2(\pi \k)}{\k} f_3 n_3 \;   \frac{f_1^2 n_1^2}{f_1^2 n_1^2 +f_2^2 n_2^2}, \\
\mu_1&=& \k_r \frac{f_2}{f_1} \sec^2 \psi - \k_\psi \left( \frac{f_2}{f_1}\right)_r \tan \psi,  \\
\mu_2 &=& \k_\theta \left( \frac{f_2}{f_1}\right)_r \tan \psi, \\
\mu_3 &=& \k_\theta  \frac{f_2}{f_1} \tan \psi.
\end{eqnarray*}

For the Taub-NUT Skyrme field, we have that 
\be \label{fiTN}  f_1 =f_2 = -\frac{r}{r+1}, \quad f_3 = \frac{r(r+2)}{(r+1)^2},\ee
 and  the integral (\ref{asympint}) becomes
\be  \label{asympintTN} \int   \om_1 \Bigr|^{\theta = \pi}_{\theta = 0}\; dr  d\psi + 
 \int   \om_2\Bigr|^{\psi = 4\pi}_{\psi = 0}\; d\theta  d r +  
\int   \om_3 \Bigr|^{r=0}_{r = \infty}\; d\theta  d\psi.
\ee

\vspace{0.2cm}

It follows that $\om_2$ is identically zero, and $\om_3 \ra 0$ as $r\ra 0$ or $r \ra \infty.$ \,
Thus the only nonzero contribution comes from the first term in (\ref{asympintTN}).  It can be shown that 
\[ \om_1 \Bigr|^{\theta = \pi}_{\theta = 0} = -2\pi \sin^2(\pi f_3) f_{3r}.  \]
This yields the topological degree (\ref{Dom}) for the Taub-NUT Skyrme field
\[D_{TN} =2,\]
which is consistent with the result in \cite{D13} obtained by counting preimages.

\vspace{1cm}

For the Atiyah-Hitchin Skyrme field, the integral (\ref{asympint}) becomes
\be \label{AH}   \int d \om =
 \int   \om_1 \Bigr|^{\theta = \pi}_{\theta = 0}\; dr d\psi + 
 \int   \om_2 \Bigr|^{\psi = 2\pi}_{\psi = 0}\; d\theta d r +  
\int   \om_3 \Bigr|^{r=\infty}_{r = \pi}\; d\theta d\psi.  \ee
However, we only know the asymptotic expressions for $f_j(r):$

For large $r,$ 
\be \label{AHlarger} f_1 =f_2 = -\frac{r}{r-2}, \quad f_3 = \frac{r(r-4)}{(r-2)^2}, \ee
and for $r$ close to $\pi,$
\be \label{AHrclosetopi} f_1 = \frac{\pi -r}{\pi} -3, \quad f_2 =\frac{\pi -r}{\pi}, \quad f_3 = \frac{r-\pi}{r+\pi}. \ee

Now it turns out that if we choose $\h \om$ in (\ref{1sthom}), the only nonzero contribution of the integral (\ref{AH}) comes 
from the first term which we are unable to integrate  as the expressions of $f_j(r)$ are not known.  \, However, we can 
proceed with the same method by choosing a different $\h \om$ as 
\be \label{2ndhom}   \h \om' = \frac{1}{2}  \left( \Psi -  \frac{1}{2} \sin 2\Psi \right) \sin \Theta \; d \Theta \wedge d \Phi,\ee
where we have put the prime in to distinguish it from $\h \om$ in (\ref{1sthom}).

Then the pullback $ \om' = {}^*\h \om'$ takes the form
\[ \om' = \frac{1}{2} \left( \frac{1}{2} \sin (2 \pi \k) - \pi \k  \right) \; \frac{f_1^2 n_1^2}{f_1^2 n_1^2 + f_2^2 n_2^2} \; d\left( \frac{f_3}{\k} \cos \theta \right)
\wedge d\left(\frac{f_2}{f_1}  \tan \psi \right).  \]
Again, with $\om' = \om_1' \, dr \wedge d \psi +\om'_2 \, d\theta \wedge d r +\om'_3 \, d\theta \wedge d \psi,$ this results in  
\[ \om_j' =  \beta' \, \mu_j',  \]
where
\vspace{-0.3cm}
 \begin{eqnarray*}
\beta' &=& \frac{1}{2} \left( \frac{1}{2} \sin (2 \pi \k) - \pi \k  \right) \; \frac{f_1^2 n_1^2}{f_1^2 n_1^2 + f_2^2 n_2^2}, \\
\mu_1' &=&   \frac{f_2}{f_1}  \left(  \frac{f_{3}}{\k} \right)_r \cos \theta  \sec^2\psi 
+ \left( \frac{f_2}{f_1} \right)_r  \frac{ \k_\psi   }{\k^2}  f_3   \cos \theta  \tan \psi,  \\
\mu_2' &=&   \left( \frac{f_2}{f_1} \right)_r \left( \frac{ \cos \theta }{\k} \right)_\theta f_3 \tan \psi,   \\
\mu_3' &=&  \frac{f_2}{f_1} \left( \frac{ \cos \theta }{\k} \right)_\theta  f_3\sec^2 \psi. 
\end{eqnarray*}
We note here that using $\h \om'$ in (\ref{2ndhom}) we consistently obtain the topological degree for the Taub-NUT Skyrme field $D_{TN} =2.$
Only that this time the integrands in the first and second terms of (\ref{asympintTN}) vanish, and only the third term gives nonzero contribution.

\vspace{0.3cm}

The same situation happens for the Atiyah-Hitchin Skyrme field, i.e. the integrands in the first and second terms of (\ref{AH}) vanish.  
Now, we can evaluate the last term using the asymptotic expressions (\ref{AHlarger}) and (\ref{AHrclosetopi}).

As  $r \ra \infty,$ one has that $(f_1, f_2, f_3) \ra (-1, -1, 1)$  and $\k \ra 1.$
 Then, 
\[ \beta' \ra - \frac{\pi}{2} \cos^2 \psi. \]
Also, it can be shown that $\k_\theta \ra 0$ as $r \ra \infty,$ thus
\[\mu_3' \ra -\sin \theta \sec^2 \psi.\]
Therefore
\[ \lim_{r \to \infty}\om_3' =  \frac{\pi}{2} \sin \theta.\]

\vspace{0.3cm}

Now as  $r \ra \pi,$ we have $(f_1, f_2, f_3) \ra (-3, 0, 0),$ which implies that $\dsl \lim_{r \ra \pi}\om_3' =  0.$
This gives the topological degree for the Atiyah-Hitchin Skyrme field
\[ D_{AH} = -1,\]
which agrees with that in \cite{D13}, obtained by the method of preimage counting, up to a sign.


\section{Relation to the $SU(\infty)$-Toda equation}  \label{sec:SUinf}
\setcounter{equation}{0}

In this section we shall investigate the Einstein-Weyl metric on the space where the Skyrme fields live. 
 This study aims to provide a link between two nonlinear problems; the non-integrable Skyrme model on one hand and the integrable
Einstein-Weyl structure on the other.  The twistor correspondence which reveals the integrability of the Einstein-Weyl structure was given by Hitchin \cite{H82a}, and all Einstein-Weyl 
spaces are governed by integrable equations (see e.g. \cite{ADM17}). In our case, it is the $SU(\infty)$-Toda equation.

Regardless of the gauge, the multi-Taub-NUT Skyrme fields constructed in Section \ref{sec:Skyrmconstr} are defined on the space of orbits ${\mc B}$
 of the Killing vector field $\dsl \frac{\p}{\p \phi}.$  There is a natural Einstein-Weyl metric on $\mc{B}$ induced by the
multi-Taub-NUT metric (\ref{multiTNsph}).    The relation between self-dual conformal structures and Einstein-Weyl structures has been established in the 
Jones-Tod correspondence \cite{JT85}, which states that any self-dual conformal structure with a conformal Killing vector field gives rise to an Einstein-Weyl structure on the 
space of orbits, and conversely given an Einstein-Weyl space one can always construct an associated self-dual conformal structure.

A multi-Taub-NUT metric is self-dual and also hyperK\"ahler, which implies Ricci-flat.  Hence it belongs to a special class, considered by 
LeBrun \cite{L91}, of scalar-flat K\"ahler metrics.\footnote{Any K\"ahler metric with vanishing scalar curvature has self-dual Weyl tensor -- in fact anti-self-dual with respect to the complex orientation.}  
In \cite{L91}, it was proved that    
 any scalar-flat K\"ahler metric with a Killing symmetry preserving the K\"ahler form can be written  
in a particular form, and is determined by solutions of the $SU(\infty)$-Toda equation 
\be 
\label{Todaeq}
u_{xx}+ u_{yy}+ \left( e^u \right)_{tt} = 0,
\ee
and its linearisation
\be \label{monopoleeq} W_{xx}+ W_{yy}+ (We^u)_{tt} = 0.  \ee

 We will see shortly that the Killing vector field $\dsl \frac{\p}{\p \phi},$ and more generally a Killing vector field of the form 
$\dsl \frac{\p}{\p \phi} + c \frac{\p}{\p \psi},$
 where $c$ is a constant,
 preserve a K\"ahler form.  Suppose we choose coordinates so that the Killing vector field considered is of the form $\dsl \frac{\p}{\p T},$ where $T$ is one of the new
coordinates.
 It follows that  the multi-Taub-NUT metric $g$ (\ref{multiTNsph}) and the K\"ahler form $\om$ can be written in the LeBrun ansatz as
\be
\label{LeBrunansatz} 
g = W h +
\frac{1}{W}(d T + \ll)^2,  \qquad  h=e^u(d x^2+d y^2)+d t^2, \ee
\be \om = W e^u d x \wedge d y +  (d\phi  + \ll) \wedge d t \label{LeBrunKahlerform}
\ee
where $\{x, y,  t\}$ are coordinates on the
space of orbits, and $(u, W),$ $\ll$  and $h$ are  functions, a one-form and a metric on
the space of orbits, respectively.    The function $u=u(x, y, t)$ necessarily satisfies the 
$SU(\infty)$-Toda equation (\ref{Todaeq}) and $W$ satisfies the so-called monopole equation (\ref{monopoleeq}).

 To see this, we note that any metric with a Killing
vector field $K=\dsl \frac{\p}{\p T}$ necessarily takes the form of the first equation in (\ref{LeBrunansatz}), where $\frac{1}{W} = g(K,K)$
and $h$ is a metric on  the three-dimensional space
of orbits of $K$.  As $K$ preserves the K\"ahler form $\om,$  the vanishing of ${\mc L}_K \om = d (K \hook \om)$ implies that
\be \label{ztilde} 
K\hook\omega=d t
\ee 
for some function $t$ on the space of orbits.  Next, we can use isothermal
coordinates $ x, y$ on the orthogonal complement of the space
spanned by $K$ and $I(K),$ together with $t,$
as local coordinates.  This will result in the form (\ref{LeBrunansatz}).
The equations (\ref{monopoleeq}) and (\ref{Todaeq}) come from
the K\"ahler condition and the 
 scalar-flat condition, respectively. The metric $h$ is the Einstein-Weyl metric in the Jones-Tod correspondence. 

Let us note here that a scalar-flat K\"ahler metric $g$ in (\ref{LeBrunansatz}) is also Ricci-flat if and only if ${u_t = kW,}$ for some constant $k$ \cite{BF82}.  
This is the case for the multi-Taub-NUT metric which is hyperK\"ahler.
	
\subsection{Metric on the space of orbits}

We shall now consider the Einstein-Weyl metric arising from the multi-Taub-NUT metric (\ref{multiTNsph}) with a Killing vector 
field of the form\footnote{The author is grateful to the anonymous referee for suggesting this more general 
one-parameter family of Killing vector fields.  As we shall see after the proof of Proposition \ref{Toda_prop}, the inclusion of parameter $c$ sheds more light
on the nature of the singularities of the Einstein-Weyl metric. } $\dsl \frac{\p}{\p \phi} + c \frac{\p}{\p \psi}.$  The metric on the space ${\mc B}$ where our Skyrme fields live corresponds to the case $c=0.$  
In the following proposition we shall give an implicit expression of the metric and identify its associated solution of the  $SU(\infty)$-Toda equation (\ref{Todaeq}). 
For convenience, in this section let us consider the multi-Taub-NUT metric in the 
cylindrical coordinates:  
\be \label{multiTNmetcylin}
g = V(d \rho^2 + \rho^2 d\phi^2 + d z^2) + V^{-1}(d\psi +\a)^2,
\ee
where we recall that $\dsl V(\rho, z)$ is a solution of the Laplace's equation 
\be\label{Laplacecylin}  \rho V_{zz} + (\rho V_\rho)_\rho  =0,  \ee
 and $\a = \h \a(\rho,z) d \phi.$
The relation $d \a = *_3 dV$ implies that  
\be \label{relValphcylin}  \h \a_z = -\rho V_\rho, \qquad \h \a_\rho = \rho V_z. \ee

\vspace{0.5cm}

\begin{prop}
\label{Toda_prop}
Let $g$ be the axially symmetric multi-Taub-NUT metric given by {\rm(\ref{multiTNmetcylin})} with 
\be \label{Vandalpha}
V = 1 + \sum_{n=1}^N \frac{1}{\sqrt{\rho^2 + (z-z_n)^2}} \quad \mbox{and} \quad
 \a = \sum_{n=1}^N \frac{z-z_n}{\sqrt{\rho^2 + (z-z_n)^2}} \, d \phi.
\ee
Then the Einstein-Weyl metric on the space of orbits of the Killing vector field
 ${K = \dsl \frac{\p}{\p \phi} +c \frac{\p}{\p \psi},}$ where $c$ is a constant, is of the form $ h=e^u(d x^2+d y^2)+d t^2,$ where $u$ is the 
solution to the $SU(\infty)$-Toda equation {\rm(\ref{Todaeq})} which is independent of $y$ and 
given implicitly by $u(x,t) = \ln(\rho^2)$ and
\begin{eqnarray}
x &=& -z + \ln \left( \frac{\rho^{N+c}}{ \dsl \prod_{n=1}^N \left(z-z_n + \sqrt{\rho^2 + (z-z_n)^2}\right)}  \right), \label{tildexim} \\
t &=&  \frac{\rho^2}{2}  + \sum_{n=1}^N \sqrt{\rho^2 + (z-z_n)^2} + c z. \label{tildezim}
\end{eqnarray}
\end{prop}

\vspace{0.5cm}

\noindent {\bf Proof} \quad  First it can be shown that the Killing symmetry generated by  ${K = \dsl \frac{\p}{\p \phi} +c \frac{\p}{\p \psi}}$ preserves a K\"ahler structure of (\ref{multiTNmetcylin}).
 The complex structure $I$ is defined by the holomorphic basis of one-forms
\[ e^1 = \frac{1}{\sqrt{V}}(d \psi + \a) + i \sqrt{V} dz, \qquad e^2 = \sqrt{V} d \rho - i \sqrt{V} \rho d \phi,\]
and the K\"ahler form given by
\be \label{Kahlerform} \om = (d \psi + \a) \wedge d z  + \rho V d\phi \wedge d \rho. \ee
It follows from (\ref{relValphcylin}) that the Lie derivative ${\mc L}_K \om = d (K \hook \om)$ vanishes.   
Then by the theorem of LeBrun, there exists a local coordinate system $\{\phi, x, y, t \}$ such
that the metric (\ref{multiTNmetcylin}) takes the form (\ref{LeBrunansatz}). 

\vspace{0.3cm}

To obtain the metric $h$ of (\ref{LeBrunansatz}), 
first we complete the square and rearrange 
(\ref{multiTNmetcylin}). This is done most conveniently in adapted coordinates in which
 $K = \dsl \frac{\p}{\p \phi} +c \frac{\p}{\p \psi} =   \frac{\p}{\p T},$ where $T$ is one of the coordinates.  

Let $\rho' = \rho,\, z'=z,\, T = \phi$ and $X = \psi - c \phi$ be the new coordinates.  For convenience, let us abuse the notation and drop the prime from 
$\rho'$ and $z'.$  Then in the adapted coordinates $(\rho, z, T, X)$
 the metric (\ref{multiTNmetcylin}) is given by
\be \label{multiTNmetadapt}
g = V(d \rho^2 + \rho^2 d T^2 + d z^2) + V^{-1}(d X +\t \a \, d T)^2,
\ee
where $\t\a := \h \a +c.$

Now completing the square and rearranging 
(\ref{multiTNmetadapt}), one obtains 
\be \label{h} h = \rho^2 d X^2 + (V^2 \rho^2 + \t \a^2) (d\rho^2 + d z^2), \ee
\[W = \frac{V}{V^2 \rho^2 +\t \a^2} \quad \mbox{and} \quad \ll = \frac{\t \a}{V^2 \rho^2 +\t \a^2} \, d X.\]
Then from (\ref{ztilde}), we have 
\be \label{tildez} d t  = \rho V d\rho  + \t \a \, d z.  \ee

The other two coordinates, $x$ and $y,$ can be chosen such that 
\be \label{tildexy} d x = \rho^{-1} \t \a \,d \rho - V d z , \qquad  d y = d X,  \ee
so that $h$ is of the form (\ref{LeBrunansatz}).  Then we have
\be \label{uimplicit} e^u = \rho^2.\ee
Equation (\ref{uimplicit}) gives the solution $u$ of the $SU(\infty)$-Toda equation implicitly.  It can be readily verified using the chain rule that
$u = \ln (\rho^2)$ indeed satisfies (\ref{Todaeq}).  By integrating (\ref{tildez}),(\ref{tildexy}), one obtains (\ref{tildexim}),(\ref{tildezim}).
\koniec

\vspace{0.3cm}

The determinant of the metric $h$ (\ref{h}) is given by $ \det h = \rho \, (V^2 \rho^2 + \t \a^2).$  Thus, $h$ may be degenerate
at $\rho = 0,$  $\rho = \infty$ or at points determined by the function $V^2 \rho^2 + \t \a^2.$     By computing the curvature invariants, 
namely the Ricci scalar and the norm of the Riemann tensor,  of  the metric (\ref{h}) with low number of centres,
 $N = 1,2, \dots, 6,$ one sees that  both invariants go to zero as $\rho \ra \infty$ in all cases, thus the metric is regular there. More importantly, 
we find that all singularities of $h$ arise from the fixed points of ${K = \dsl \frac{\p}{\p \phi} +c \frac{\p}{\p \psi}},$ i.e. where the norm $|K|$ vanishes, which depend on the 
number of centres of the multi-Taub-NUT metric and the value of $c.$

For $N=1,$ $K$ has only one fixed point at the centre  for a general value of $c,$ the exception being $c = \pm 1.$  This can be
 seen from  the norm squared  $|K|^2 =(V^2 \rho^2 + \t \a^2)V^{-1},$ which vanishes when $\rho = 0$ and $\t \a = 0.$  Now imposing $\rho =0$ and
without loss of generality  setting the centre at the origin, we have $\dsl \t \a = \frac{z}{\sqrt{z^2}} + c.$     The values $c = \pm 1$ are thus singled out  as the
 term $\dsl \frac{z}{\sqrt{z^2}}$ can only take values $\pm 1.$  The function $\t \a$ then vanishes when $z< 0$ ($z> 0$) for $c=1$ ($c=-1$).  
In these two cases, the centre is no longer an isolated fixed point, but is part of a two-dimensional surface of fixed points.
 The fixed points of $K$ lead to the singularities of the Einstein-Weyl metric $h$, where
the Ricci scalar and the norm of the Riemann tensor blow up. These singularities lie on the  $\rho = 0$ boundary of the $3$-dimensional quotient space.

The same behaviour happens for other $N,$ where in general the only singularities of the metric $h$ correspond to the centres of the
multi-Taub-NUT metric, which are isolated fixed points - nuts - of the isometry $K.$  In this case, $h$ is regular everywhere on the quotient space except 
for $N$ isolated points on the $\rho = 0$ boundary.

Exceptional cases occur for some values of $c;$ the particular values are
 determined by the number of centres, $N.$ \, From the expression of $\t \a,$ one finds that the exceptional values of $c$ are $\pm N, \pm (N-2), \pm (N-4), \dots, \pm 1 \, (0)$ 
for odd (even) $N.$ \, For each of these values of $c,$ there exists exactly one $2$-dimensional surface of fixed points, either connecting a pair of centres or, in the case $c = \pm N,$ 
starting from a centre at an end of the centre arrangement towards the direction $z \ra \pm \infty.$  The rest of the centres remain nuts.\, The metric $h$ is singular at the corresponding points on the space of orbits of the isometry.

Let us note here that although the locus of fixed points of the isometry $K$ can be found for all $N,$ we have checked the singularity of $h$ explicitly only for the case $N=1,2, \dots, 6,$ as mentioned previously.   From the form of $V$ and  $\a$ (\ref{Vandalpha}) governing the metric $h$ (\ref{h}),  we expect  that the singularities only come from the fixed points of $K,$ and $h$ is regular everywhere else for any 
$N.$  However, we are still unable to obtain
a tractable general expression for the curvature invariants for any $N$  which would prove the claim.

\vspace{0.3cm}

The Skyrme fields constructed in Section \ref{sec:Skyrmconstr} live on the space of orbits ${\mc B}$
 of the Killing vector field $K = \dsl \frac{\p}{\p \phi},$ i.e. the case $c=0.$  From the above discussion, it follows that for the odd $N$ the Einstein-Weyl metric $h$  
 has only $N$ isolated singularities on the $\rho = 0$ boundary of ${\mc B}.$  As these arise from the nut fixed points of the vector field $K$  used to obtain the 
quotient ${\mc B},$ they are not part of the quotient.  Thus we conclude that the metric is regular on ${\mc B}.$

For even $N,$  the curvature invariants 
are also singular on the part of the $\rho = 0$ boundary which connects the two points corresponding to the middle two centres. For example, for $N=4,$
suppose the centres are located at $z= z_n$ with $z_1 < z_2 < z_3 < z_4$ on the $z$-axis in $\R^3$, then the invariants are singular for $z_2 < z < z_3$ on 
the $\rho = 0$ boundary of ${\mc B}.$  

We note here that, for the purpose of the Skyrme field  construction,
 one can avoid this non-isolated singularity for even $N$ by calculating the holonomy of the spin connection along  the orbits of a  Killing vector field 
${K = \dsl \frac{\p}{\p \phi} +c \frac{\p}{\p \psi}}$ with some `good' values of $c$ (e.g. $c \ne 0, \pm 2, \pm 4$ for $N=4$)  instead.

\vspace{0.3cm}


\subsection{Limit $N \ra \infty$}

\vspace{0.2cm}

As $N$ becomes large, the function $V$ of the multi-Taub-NUT metric (\ref{multiTNmetcylin}) approaches the
Ooguri-Vafa limit \cite{OV96}, such that when $\rho \ra \infty,$ $V$ can be approximated by
\[  V'  = -\ln(\rho^2). \]
Then, using the same procedure as in the proof of Proposition \ref{Toda_prop}, one obtains the  Einstein-Weyl metric $h$ on 
the space of orbits of the vector field $K = \dsl \frac{\p}{\p \phi}+ c\frac{\p}{\p \psi},$ of the form $ h=e^u(d x^2+d y^2)+d t^2,$ where
the solution to the $SU(\infty)$-Toda equation $u$ is given implicitly by
 $u(x,t) = \ln(\rho^2)$ and
\be \label{OVform}
x = z \ln(\rho^2) + c \ln \rho, \qquad
t =  z^2 + c \, z - \frac{1}{2} \rho^2 (\ln(\rho^2) -1).
\ee
The relations (\ref{OVform}) show that $u(x,t)$ satisfies 
\[ \frac{e^u}{2}  u^2 (u-1)  + u^2 \left(t+ \frac{c^2}{4} \right) -x^2 = 0,\]
which implies that $u$ is a solution which is constant on a cylinder $\dsl k_1 \left(t + \frac{c^2}{4} \right) -x^2 = k_2,$ 
where $k_1, k_2$ are constants.
  
We note here that solutions of the $SU(\infty)$-Toda equation constant on surfaces have been studied previously 
in \cite{T95} and \cite{DP11}, where the surfaces are central ellipsoids and planes, respectively.


\subsection{More explicit expressions}

\vspace{0.2cm}

To obtain 
an explicit expression for the solution $u(x, t)$  in Proposition \ref{Toda_prop}, one needs to know $\rho(x, t)$ explicitly. 
In principle, this can be achieved by inverting the expressions  {(\ref{tildexim}),(\ref{tildezim})}.
However, it turns out to be impossible  even for the case $N=1$ of the Taub-NUT metric.  Therefore, to obtain a 
more explicit description of the corresponding solution to the $SU(\infty)$-Toda equation and thus the Einstein-Weyl metric, 
we proceed analogously  to \cite{W90} as we shall describe below.

First, we note that the solution $u$ in Proposition \ref{Toda_prop} is
 independent of $y.$  Such a solution belongs 
 to the class of solutions considered by Ward in \cite{W90}.   There, Ward gave a prescription of how one can obtain a 
 solution of the $SU(\infty)$-Toda equation from an axially symmetric solution of the Laplace's equation.\footnote{The procedure in \cite{W90} was in fact formulated in the $2+1$ dimension, with the wave equation in place of the Laplace's equation.  However, it can be readily adapted to the Euclidean signature.}
We shall apply his procedure to the solutions of the Laplace's equation governing the multi-Taub-NUT metrics of Proposition \ref{Toda_prop}.
We will show that this method yields an explicit expression for $u$ in the Taub-NUT $(N=1)$ case, and that $u$ is 
determined by a root of a quintic polynomial for $N=2.$ 

\vspace{0.3cm}

The multi-Taub-NUT metric (\ref{multiTNmetcylin}) is governed by the axially symmetric solution $V$  of the Laplace's equation
\be \label{axisymLaplace}  \rho V_{zz} + (\rho  V_\rho)_\rho  =0,  \ee
given in (\ref{Vandalpha}). 
One notes that  the function $\wh V$
\be \label{whV} \wh V =  \sum_{n=1}^N \ln \left[ (z-z_n) + \sqrt{\rho^2 + (z-z_n)^2} \right], \ee
 satisfying
 $ V = 1 + \wh V_z,$
is also a solution of (\ref{axisymLaplace}).
Now, define new variables $(\xi, \tau)$  by  
\be \label{defxt} \xi = \wh V_z,\qquad  \tau = \frac{\h \a + c}{2},  \ee
where we recall that    the function $\h \a$ is given in (\ref{Vandalpha}) (notice that $\h \a = N - \rho \wh V_\rho$), and that $\h \a + c,$ 
denoted previously as $\t \a,$ appears in the expression for the Einstein-Weyl metric (\ref{h}) on the space of orbits of the Killing symmetry generated by
$K = \dsl \frac{\p}{\p \phi}+ c\frac{\p}{\p \psi}.$  The choice of $\tau$  to include the constant $c$ is for the convenience in writing the metric (\ref{h}) in these new coordinates.

Inverting (\ref{defxt}) to obtain $\rho(\xi,\tau),$  it follows that if we let 
\be \label{u} u = \ln\left(\frac{\rho^2}{4}\right),\ee then $u$ satisfies the 
reduced $SU(\infty)$-Toda equation
\be \label{xtToda}   u_{\xi\xi} + (e^u)_{\tau\tau} =0.  \ee

\vspace{0.3cm}

Our aim now is to obtain an expression, as close to being explicit as possible, for $\rho(\xi, \tau)$ for $\wh V$ in (\ref{whV}).  The change of variables (\ref{defxt}) gives
\be \label{eqnforxt}
\xi = \sum_{n=1}^N y_n, \hspace{2cm}
 2\tau = z \xi - \sum_{n=1}^N  z_n \, y_n  + c, 
\ee
where $\dsl y_n = \frac{1}{\sqrt{\rho^2 +(z-z_n)^2}}.$

\vspace{0.5cm}

In principle, one could use the equations (\ref{eqnforxt}) together with the expression for $y_n(\rho, z)$ to obtain
$\rho^2(\xi,\tau).$  The idea is best illustrated through the following examples of the Taub-NUT metric $(N=1)$ and the multi-Taub-NUT metrics with $N=2$
and $N=3.$   

\vspace{0.3cm}

First, let us consider the Taub-NUT metric, where $N=1,$ $z_1 =0.$  The equations (\ref{eqnforxt}) become
\[ \xi= y_1, \hspace{2cm} 2\tau = z \xi +  c. \]
Here we simply write $z=(2\tau -  c)/\xi$ and, from $\dsl y_1^2 = \frac{1}{\rho^2 +z^2},$ obtain
\[\rho^2 = \frac{1}{\xi^2} - z^2 = \frac{1-4 \, (\tau - c/2)^2}{\xi^2}.\]
Thus the corresponding solution of the $SU(\infty)$-Toda equation, $\dsl u = \ln\left(\frac{\rho^2}{4}\right),$ is given explicitly by 
\[ u = \ln\left(\frac{1/4 -\, (\tau - c/2)^2}{\xi^2}\right).\]

\vspace{0.5cm}

Next, for the multi-Taub-NUT $N=2,$  let $z_1 = 0$ and $z_2 = 1$ for simplicity.  We now have
\[ \xi = y_1 + y_2, \hspace{2cm} 2\tau = z\xi -y_2 + c. \]
This is a linear system from which we can easily write $y_1, y_2$ as functions of $\xi, \tau$ and $z:$
\[ y_1 = (1-z)\xi + 2\tau - c,  \hspace{2cm}  y_2 = z\xi - (2\tau - c).\]
Now, using the expressions of $y_1$ and $y_2,$ one obtains
\be \label{N2eqn}  \rho^2 + z^2 = \frac{1}{(\,(1-z)\xi + 2\tau - c \,)^2}, \hspace{2cm} \rho^2 + (z-1)^2 = \frac{1}{(z \xi- (2\tau - c))^2}. \ee
The difference of the two equations in (\ref{N2eqn}) gives a quintic polynomial equation for $z$ in terms of $\xi, \tau:$

\[ \label{N2poly} (\,(1-z)\xi + 2\tau - c \,)^2 (z \xi- (2\tau - c))^2 (2z-1) = 2\xi (z\xi- (2\tau - c) ) - \xi^2. \]
The root of the polynomial which satisfies the system (\ref{N2eqn}) will give the desire expression for $z(\xi,\tau),$ which in turn
yields $\dsl u(\xi,\tau) = \ln\left(\frac{\rho^2}{4}\right),$ via 
\[   \rho^2 = \frac{1}{ (\,(1-z)\xi + 2\tau - c \,)^2 - z^2}. \]

\vspace{0.5cm}

Lastly, to see how the computation develops  as $N$ becomes larger, let us consider the case of the 
multi-Taub-NUT metric with $N=3.$  For simplicity, let $z_1 = 0, z_2 =1$ and $z_3 = -1.$  Then  (\ref{eqnforxt}) becomes
\be \label{N3xteqn}   \xi= y_1 +y_2 + y_3, \hspace{2cm}  2\tau = z \xi -y_2 + y_3 + c. \ee

Similar to the $N=2$ case, one can start by writing $\dsl y_3 = \frac{y_2}{\sqrt{1+4z \, y_2^2}}.$ 
The system (\ref{N3xteqn}) is no longer a linear system of equations for $y_1, y_2.$  Thus, to
get expressions for $y_1(z, \xi, \tau), \, y_2(z, \xi, \tau),$ one needs to solve a polynomial equation.  
With $y_1(z,\xi,\tau), \,y_2(z,\xi,\tau),$ one can then proceed similarly to the case $N=2$ to get $z(\xi,\tau)$ from 
\[  \rho^2 + z^2 = \frac{1}{y_1^2(z,\xi,\tau)}, \hspace{2cm} \rho^2 + (z-1)^2 = \frac{1}{y_2^2(z,\xi,\tau)}. \]
The difference of the two equations above will give a high order polynomial for $z.$  The root $z(\xi,\tau)$ would then yield $\rho^2(\xi,\tau)$
via $\dsl \rho^2 = \frac{1}{y_1^2 - z^2}$ as before.

\vspace{1cm}

Let us conclude this section by writing down the Einstein-Weyl metric (\ref{h}),
 \[ h = \rho^2 d X^2 + (V^2 \rho^2 + \t \a^2) (d\rho^2 + d z^2), \]
in the coordinates $(\xi,y, \tau)$ on the space of orbits.  \; From
\[ \xi = V-1, \qquad \quad  y = X,  \qquad \quad \tau = \frac{\t \a}{2},  \]
and letting $e^u = \rho^2/ 4,$  
the metric $h$ becomes
\[h/4 = e^u\, (\mathcal{H}\, d\xi^2 + dy^2) + \mathcal{H}\, d\tau^2,\]
where $\mathcal{H}(\xi,\tau)$ is a function given by 
\[ \mathcal{H} = \frac{ e^u(1+\xi)^2 + \tau^2}{e^u (\xi_\rho^2 + \xi_z^2)},\]
and $\xi_\rho, \xi_z$ denote $\dsl \frac{\p \xi}{\p \rho}$ and $\dsl \frac{\p \xi}{\p z},$ respectively.

\vspace{0.5cm}

The difficulty in writing $h$ explicitly 
comes down to writing $\xi_\rho, \xi_z$ as functions of $(\xi, \tau).$
A direct computation shows that this requires an expression for $z(\xi, \tau),$ which we have obtained only
for the Taub-NUT case, where
\[\mathcal{H} = \frac{e^u \, (1+\xi)^2 + \tau^2}{e^u \, \xi^4}. \]

\vspace{1cm}


\section{Further remarks}  \label{sec:Discuss}
\setcounter{equation}{0}

Let us recall that the Euler-Lagrange equation of the Skyrme model is not integrable and solutions have only been found numerically. 
There are several known methods of constructing analytic approximations, which we also call Skyrme fields in this paper, to the true static solutions.  
Some examples are  the product ansatz \cite{Sch94}, the rational map ansatz \cite{HMS98} and the construction of Skyrme fields  from instantons \cite{AM89}, which is the main focus of this paper.
 
Recall that static solutions to the Skyrme equation are critical points of the Skyrme energy functional.  Therefore, one desires a field profile which is close to a critical point.
In \cite{AM89},   generalisation 
of 't Hooft's instantons \cite{JNR77} were shown to generate Skyrme fields on $\R^3,$ some of which have energy very close to the true solutions.  
  For example, the one-pole 't Hooft's instanton gives a spherically symmetric Skyrme field with a parameter.  This turns out to be a good approximation to the 
minimum energy Skyrmion solution with the topological degree $D=1$ - the  minimum energy of the profile, obtained by adjusting the parameter, is less than $1\%$ above the energy of the true (numerical) Skyrmion.  

\vspace{0.3cm}

The energy functional derived from the standard Lagrangian of the Skyrme model is given by
\be \label{energy} E = \int \left( -\frac{1}{2} \mbox{Tr}(R_a R^a) -  \frac{1}{16}  \mbox{Tr}([R_a, R_b][R^a, R^b]) \right) dV_3,\ee
where $d V_3$  denotes the volume form of the three dimensional space,  $R_a =U^{-1} \p_a U,$ with $\p_a U$ denoting the derivative with respect to 
each of the three coordinates, and the Einstein summation convection is used.

In our case, to see how well the multi-Taub-NUT Skyrme fields (\ref{UmultiTN}) approximate the solutions to the Skyrme equation on  ${\mc B},$
one needs to consider the energy functional (\ref{energy}) on  ${\mc B}.$  

As a start, let us focus on the Taub-NUT Skyrme field $N=1$, where the Skyrme field
 (\ref{UmultiTN}) simplifies to 
\be \label{Uansatz} U(r, \theta, \psi) = \exp \left( -i \,\pi \, f_{j}(r) n_j \, \tau_j\right),\ee
which can be written as (\ref{Utau}) with $f_j$ given by $(\ref{fiTN}).$  

\vspace{0.5cm}

To compare the energy of this field with a true solution, one needs to write down a Skyrme equation on ${\mc B}$ and then solve it numerically.  Static solutions satisfy the 
Euler-Lagrange equation of the energy functional.  Therefore we first write down the energy functional (\ref{energy}) for an
 ansatz of the form (\ref{Uansatz}) with $f:=f_1 = f_2$ and $g:=f_3,$ which is a slight extension of the Hedgehog ansatz, where $f_1=f_2 =f_3.$  

In hindsight, let us assume the metric  $h_{\mc B}$ on ${\mc B}$  to be used in the energy functional  to be of the form
\be \label{hBform}  h_{\mc B} = F \, (dr^2 + r^2 d \theta^2) + G  \, d \psi^2, \ee 
where $(r, \theta, \psi)$ are the coordinates of Section \ref{sec:Skyrmconstr} in which our ansatz (\ref{Uansatz}) are written, and $F, G$ are some functions  

Substituting the ansatz in (\ref{energy}) and using the metric (\ref{hBform}) 
 yields quite a complicated expression
\begin{eqnarray}   E = \int_{\mc B}   &\Big\{& F^{-1}K(r,\theta) \, +\, \sin^2(\kappa) \, \Big(  F^{-1}N(r,\theta) \, + \, \frac{(F^{-1})^2}{r^2} L(r, \theta)
\, + \, G^{-1} \frac{f^2}{\kappa^2} \sin^2\theta \, (1+  F^{-1} K(r,\theta) ) \Big)   \nonumber \\ 
&+& \sin^4(\kappa)  \, F^{-1}  \, G^{-1} \,  \frac{f^2}{\kappa^2} \sin^2\theta  \,
N(r,\theta) \; \Big\} \; \sqrt{\det h_{\mc B}} \; dr \,  d\theta \, d \psi, \label{energyfn}
\end{eqnarray}
where 
\begin{eqnarray*}
K(r,\theta) &=& \kappa_r^2 + \frac{\kappa_\theta^2}{r^2} \, = \,
\frac{1}{\kappa^2} \left( (f f' \sin^2 \theta +  g g' \cos^2 \theta)^2 + \frac{(f^2 - g^2)^2}{r^2}  \sin^2 \theta \cos^2 \theta   \right) \\
N(r,\theta) &=& |{\bf \h n}_r|^2 +\frac{|{\bf \h n}_\theta|^2}{r^2}  \, = \, \frac{1}{\kappa^4} \left( (f' g - f g')^2 \sin^2 \theta \cos^2 \theta  +\frac{f^2 g^2}{r^2}  \right) \\
L(r,\theta) &=& \kappa_r^2  \, |{\bf \h n}_\theta|^2  +   \kappa_\theta^2 \, |{\bf \h n}_r|^2  -  2 \,\kappa_r \, \kappa_\theta\, {\bf \h n}_r \cdot {\bf \h n}_\theta  \, = \,
\frac{(f'g \sin^2 \theta + f g' \cos^2 \theta)^2}{\kappa^2},
\end{eqnarray*}
and we recall from Section \ref{sec:TNAH} that, with  $f:=f_1 = f_2$ and $g:=f_3,$  we have \\$\kappa = \sqrt{f^2 \sin^2 \theta + g^2 \cos^2 \theta}$  
and  $\dsl {\bf \h n} = \frac{1}{\kappa} \left( f\sin \theta \cos \psi, f \sin \theta \sin \psi,  g  \cos \theta \right).$  Also, ${\bf \h n}_r$ and ${\bf \h n}_\theta$ denote the 
derivatives of the vectors.

\vspace{0.5cm}

Now, ${\mc B}$ has a natural Einstein-Weyl metric given by (\ref{h}) (with $c =0$), which for the Taub-NUT case can be written in $(r, \theta, \psi)$ coordinates as
\be \label{hsph} h= (\cos^2\theta + (r+1)^2 \sin^2 \theta) (dr^2 + r^2 d \theta^2) + r^2 \sin^2  \theta d \psi^2.\ee
However, using the metric (\ref{hsph}) in (\ref{energyfn}) and performing numerical integration gives unbounded energy for the Taub-NUT Skyrme field.  This is due to the fact that
the integrand of (\ref{energyfn}) blows up as $r \ra \infty.$ \, Note, however, that the integrand  has a zero limit as $r \ra 0,$ i.e. approaching the nut. 

\vspace{0.5cm}

One way to proceed is to  use another metric $h_{\mc B}$ on ${\mc B}.$   One could look for a conformal factor $\Lambda$ such that the metric 
$h_{\mc B} = \Lambda \, h,$ 
$h$ given in (\ref{hsph}), yields a finite energy.  However, so far we have failed to find or prove an existence of such a $\Lambda.$  It might be that a metric $h_{\mc B}$ in a different conformal 
class is needed. This  investigation is left for future work.

In principle, once a  metric (\ref{hBform}) is chosen, one then needs to derive the  Euler-Lagrange equation of the energy functional and solve it numerically to get a true static solution.
 Note that care is needed when imposing the boundary condition for the Euler-Lagrange equation, so that one obtains a 
solution with the same topological degree as the approximate Skyrme field.

A final remark is that in order to find a Skyrme field which best approximates a true solution, one may need to reinstate a parameter in the harmonic function 
$V$ defining the multi-Taub-NUT metric (\ref{multiTNmet}), which has been set to $1$ for simplicity.
  The parameter can then be varied to achieve the minimum energy for the Skyrme field.


\section*{Acknowledgements}
I wish to thank Maciej Dunajski, Nicholas Manton, Teparksorn Pengpan and the anonymous referees for valuable comments and suggestions. I am also grateful to the
London Mathematical Society for supporting my visit to Durham, where this work was completed.  This
research is funded by the Thailand Research Fund, under the grant number TRG5880210.



\end{document}